# A discrete random model describing bedrock profile abrasion.

András A. Sipos[1], Gábor Domokos[1], Andrew Wilson[2] and Niels Hovius[3]

[1] Department of Mechanics, Materials and Structures, Budapest University of Technology and Economics, Hungary, siposa@szt.bme.hu
[2] Stratigraphy Group, University of Liverpool, UK
[3] Department of Earth Sciences, University of Cambridge, UK

**Abstract**

We use a simple, collision-based, discrete, random abrasion model to compute the profiles for the stoss faces in a bedrock abrasion process. The model is the discrete equivalent of the generalized version of a classical, collision based model of abrasion. Three control parameters (which describe the average size of the colliding objects, the expected direction of the impacts and the average volume removed from the body due to one collision) are sufficient for realistic predictions.

Our computations show the robust emergence of steady state shapes, both the geometry and the time evolution of which shows good quantitative agreement with laboratory experiments.

# 1. Introduction

Abrasion of bedrock by impact of moving particles is a dominant process shaping obstacles in river channels (Sklar and Dietrich 2001) and persistent wind (Laity and Bridges 2009). Our goal is to identify and predict the geometry of obstacle profiles created by this process.

Origin and evolution of shapes is a widely discussed question in geological literature. Models of topographic evolution of elements of the natural landscape aim to explain morphological changes caused by erosion observed in particular geomorphological locations. These models often apply statistical relations with several empirical and place-dependent parameters (Montgomery 2003; Tomkin 2009) or identify the key components of the physical process first (Carter and Anderson 2006; Roering, Kirchner and Dietrich 2001; Smith, Merchant and Birnir 2000). However, due to the complexity, and large number of unknown parameters involved in the problem, even the latter approaches are difficult to apply as predictive models. In case of smaller objects (e.g. pebbles or rocks) experiments beyond field observations play a crucial role in understating the evolutional process (Laity and Bridges 2009; Phillips and Lutz 2008; Schoewe 1932) here a model based on physical hypotheses can be verified in a more reliable manner (Ahnert 1987). For example, the classical model of impact abrasion in the form of a geometrical partial differential equation was given by Firey (1974). In his approach abrasion is modeled as a sequence of collisions leading to material loss from the abraded object $P$. Firey assumed uniformly distributed collisions with an infinite plane $G$ (Fig. 1(a)) leading to surface wear with speed proportional to (Gaussian) curvature:

$$\dot{\mathbf{v}} = \delta f \rho(\alpha) \cdot \mathbf{n}_\alpha, \qquad (1)$$

where $\mathbf{v}$ is the position vector of a point on the perimeter with (time-dependent) normal direction $\mathbf{n}_\alpha$, $\rho(\alpha)$ is the Gaussian curvature of the perimeter, and $f$ is a constant. The average area $\delta$ lost by the impacted object is assumed to be independent of the location of the collision, dot denotes derivative with respect to time $t$.

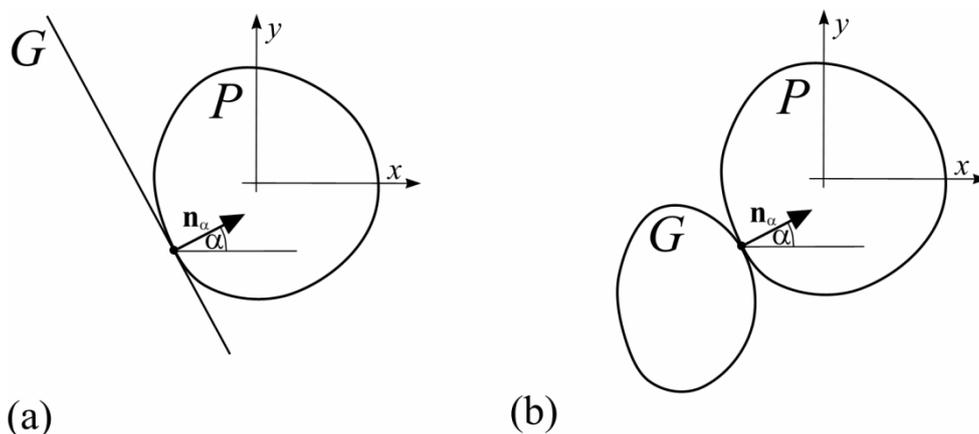

**Fig. 1** Notations for Firey's model (a) and the generalized model with a finite sized abrader (b)

Firey's model was generalized in (Domokos, Sipos and Várkonyi 2009). The two dimensional version (Fig. 1(b)) of the new partial differential equation – beyond Firey's model – includes one additional term, modelling the abrasion by infinitesimal particles, but still with uniform spatial distribution

$$\dot{\mathbf{v}} = c[\bar{p} \cdot \bar{\rho}(\alpha) + 1] \cdot \mathbf{n}_\alpha, \qquad (2)$$

where $c$ is a constant (in three dimension, the full partial differential equation consists of three terms). The scale-free parameter $\bar{p}$ represents the relative exposed perimeter of nearby stones and $\bar{\rho}(\alpha)$ is the scale-free curvature, depending only on shape but not on size.

This simple model explains the evolution of an abraded object in a reliable manner as long as the probability of collision from all directions is equal. The three dimensional version of the model was applied successfully to describe the shape evolution of asteroids (Domokos et al. 2009).

Here we argue that this simple, collision-based model can adequately describe the evolution of bedrock obstacles under a steady sediment-laden stream, such as found in rivers incising rock substrates. In this case, the river flow direction defines the dominant direction of abrasive particle movement, so the original assumption about the uniform probability of impact direction must be abandoned. Moreover, where a sediment cover is present on the channel bed, the buried portion of a bedrock obstacle surface is not exposed to any abrasion. To accommodate these fundamental features of the process we applied a discrete version of the PDE model in (Domokos, Sipos and Várkonyi 2009) and compared the simulations to field observations in Taiwanese bedrock rivers and to experimental results, presented in (Schoewe 1932; Wilson 2009; Wilson and Hovius 2010). The good agreement between the experimental data and the numerical results confirm the key role of the bedload in the erosion of bedrock bedforms.

Another advantage of our discrete approach is that it can account for finite fragmentation, a phenomenon not captured by the partial differential equation (Eq. (2)) which assumes that infinitesimal amounts of material are abraded in continuous time. Since abrasion is considered as an effect from series of discrete collisions, an approach which takes finite fragmentation into account can be regarded as more realistic.

Our simulations predict the robust emergence of stationary profiles. This agrees both with experiments and field observations (Wilson 2009; Wilson and Hovius 2010). Besides the shape and steady-state character we also found good agreement between the time evolution of simulated and observed profiles. In addition, our simulations predict that large grains traveling as bedload in frequent contact with the channel bed create upstream facing convex surfaces, which are observed on obstacles in bedrock river channels (Tomkin 2009), whereas small grains create polygonal profiles with flat faces and sharp edges, as found on ventifacts (Laity and Bridges 2009).

In Section 2 we describe our model in details, in Section 3 we compare simulations to experiments from the literature and in Section 4 we draw conclusions.

## 2. Description of the model

We assume that part $P$ (eroding profile) of the surface of an obstacle of height $h$ is impacted by grains $G$ of diameter $d$, the incoming direction of which is lognormally distributed about an expected value $\alpha$, with standard deviation $\sigma_1$ (Fig. 2). (We picked lognormal as one natural distribution with positive sign since due to gravity incoming particles almost always arrive on a downward slope.) Both colliding objects are represented in 2D by polygons consisting of vertices and edges. Since vertex-vertex and edge-edge collisions have zero probability, there

are two possible outcomes, corresponding to the two terms in the partial differential equation (Eq. (2)):

(A) collision between a vertex of $P$ and an edge of $G$ and
(B) collision between an edge of $P$ and a vertex of $G$,

with associated probability distribution $p(A)=p_0$, $p(B)=1-p_0$. The value of $p_0$ depends on the (averaged) relative size of the colliding objects. For abraders with increasing size (i.e., $d \gg h$), the solution converges to the abrasion by an infinite plane, (Firey's original model). In our discrete approach this limit can represented by the sequence of pure type (A) events, that is $p_0=1$. In the other limit ($d \ll h$, sand blasting) the discrete approach contains only type (B) events, that is $p_0=0$.

In each event a random amount with lognormal distribution (expected value $\Delta$, standard deviation $\sigma_2$) of volume (area) is abraded from the profile $P$. (Here, similarly to α, we choose lognormal as a natural distribution of a non-negative quantity). In case of event (A) we pick one value of the random variable α and select the $i^{th}$ vertex on the abraded profile $P$ such that the normal directions $\alpha_i$ and $\alpha_{i+1}$ of the adjoining faces are bracketing α, i.e. we have $\alpha_i < \alpha < \alpha_{i+1}$. Since the profile is convex, this choice is unique. Subsequently, the profile is chopped off at that vertex, creating two new near-by vertices (Fig. 2). In case of event (B) a randomly selected edge along the profile retreats parallel to itself. The random selection is weighted by the length of the edges: the probability of an edge to be sampled is proportional to its length.

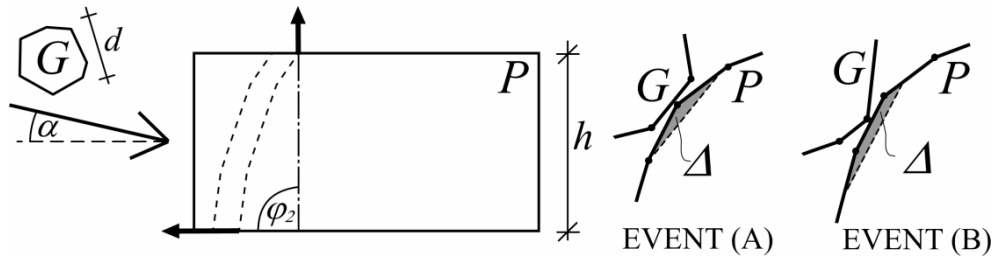

**Fig. 2** An originally rectangular profile $P$ with height $h$ is abraded by grains ($G$, with an average diameter $d$) from the incoming direction which has an expected inclination α. The abrasion of profile $P$ is modelled by the random sequence of events (A) and (B). In each step area $\Delta$ is removed from the profile $P$. The shape evolution due to the abrasion is indicated by dashed curves in the profile $P$. The aperture of the abraded profile (angle between the end-normals) is approximately $\varphi_2 \cong \pi/2$.

Observe that in this model the exact geometry of the impacting grains does not enter into the computation, we only assume that they either hit a vertex of $P$ or an edge of $P$. One collision in the model is an abstract event and it represents the averaged effect of many small hits. Intensive wear at sharp vertices and long edges is realized in the model by the higher probability of hits at these locations, due to their higher exposure. The erosion process is modelled by a randomly sampled sequence from the abstract events (A) and (B), based on the above distribution, controlled by the parameters $α, p_0, \sigma_1, \sigma_2$ and $\Delta$. The latter three do not influence the geometry of the profile: $\Delta$ plays (within certain limits) the role of time, and since the flow of impacting grains is almost unidirectional, and the grain sizes are almost constant, we assume $\sigma_1 \ll α$, $\sigma_2 \ll \Delta$, implying that neither $\sigma_1$ nor $\sigma_2$ play an independent important role. Thus the main control parameters remain $α$ and $p_0$.

Our averaged model has the advantage that its main control parameters have a direct physical interpretation. Based on our assumptions small $p_0$ corresponds to small grain size (sand blasting) whereas large $p_0$ corresponds to large grain size. We can define the relative exposure $\bar{p}$ of the impacting grain as

$$\bar{p} = \frac{p_1 \varphi_2}{p_2 \varphi_1}, \tag{3}$$

where $p_1, p_2$ and $\varphi_1, \varphi_2$ are the exposed lengths and the apertures of the grain and profile, respectively. Apertures measure the angle between end-normals, e.g. for all closed curves the aperture is $2\pi$. We use the notations of Figure 2. We assume circular grains ($p_1 \cong d\pi$) and consider the profile's exposed length to be approximately equal to its height ($p_2 \cong h$) while its aperture to be approximately the angle between the normals at the base and the top ($\varphi_2 \cong \pi/2$). Since the abraded surface of the incoming grains is always a closed curve ($\varphi_1 = 2\pi$) we can now estimate the relative exposure as

$$\bar{p} = \frac{p_1 \varphi_2}{p_2 \varphi_1} \cong \frac{d\pi(\pi/2)}{2\pi h} = 0.785 \frac{d}{h}. \tag{4}$$

The relative exposure (thus scaled $0 < \bar{p} < \infty$) can be expressed according to (Domokos, Sipos and Várkonyi 2009) in terms of the probability $p_0$.

$$\bar{p} = \frac{p_0}{1 - p_0} \quad \text{and} \quad p_0 = \frac{\bar{p}}{1 + \bar{p}}. \tag{5}$$

We remark that in case of closed curves the relative exposure is simply equal to the ratio of the diameters.

## 3. Comparison to experimental results

Bedrock river channels (Turowski et al. 2008) dominate many erosional landscapes, but it is hard to track the evolution of obstacles in these channels, and impossible to isolate grain size effects on their abrasion. This can be done in an annular, recirculating flume capable of generating hydrodynamic conditions as in mountain river floods (Attal and Lave' 2006). In an experiment with this flume (Wilson 2009), single cuboid marble blocks, 10,0 cm tall and 20 cm long, and spanning the 26 cm width of the flume, were fixed on the channel base as a flow obstruction. The flume was loaded with sorted limestone pebbles with a known starting weight and grain size range to act as bedload abrasion tools, while water discharge was held constant to produce a flow speed of ~3ms$^{-1}$. Flume conditions preserved the approximate proportionality of obstacle height, bedload size and flow depth observed in natural bedrock streams. Under these flow conditions, the limestone pebbles moved by rolling and low saltations in a layer of one to two grain diameters depth along the flume base. Particles moved up the stoss surface of an obstacle in one or several short hops, and launched off this face clearing the remainder of the obstacle. Resultant erosion of bedrock obstacles was measured using three dimensional laser scanning at intervals throughout experiment runs that lasted 400-690 minutes. In a blank run without mobile sediment, no measurable erosion occurred, and all measured erosion can be attributed to bedload impact abrasion. Abrasion of obstacles was dominantly on the stoss surface whose initially square cross section evolved to an unpstream facing convex surface (The lower 20 mm of the stoss surface were prone to edge chipping, and have been eliminated from consideration. Rates of stoss surface retreat (horizontal erosion) were initially highly variable, peaking near the top of the obstruction, and decreasing systematically towards the base. This distribution of erosion gave rise to a progressive reclining and convex-up rounding of the stoss side. Over time horizontal erosion converged to a common rate at all heights above the flume base. Thus, the obstacle stoss faces achieved a time-independent form that advanced downstream into the obstacles, the top and lee side of which underwent very little erosion by comparison. These attributes are shared with bedrock obstacles in the Liwu River, Taiwan, that have been studied extensively (Harthshorn et al 2002).

We have set up numerical simulation applying our model to match the flume experiment. Hence, the abraded profile $P$ is rectangular at the outset, with height $h=80$ mm. All simulations were run to steady state. Final profile shapes clearly reflect the dominant abrasion mechanism. Abrasion of edges by type (A) events involving large $p_0$ (large grains) results in smooth, rounded profiles consisting of many edges connected by short vertices, progressively less inclined with increasing height above the base of the obstacle (Fig. 4 (c)). Abrasion of vertices by type (B) events involving small $p_0$ (small grains) gives rise to long vertices joined by dominant edges (Fig. 4 (a)). In the flume experiment the size of moving grains was relatively large (60 mm < $d$ < 80 mm) compared to the height of profile ($h=80$mm). Substituting this data into Eq. (4) and Eq. (5) yields the relevant range of the control parameter as $0.37 < p_0 < 0.44$. The angle $\alpha$ defines the normal of the impacting surface in case of an A-type abstract event and it is strongly coupled with the average slope $\beta$ of the profile (Fig. 3). Based on numerical simulations, for the given range of $p_0$ we found

$$\alpha \sim 58.82 - 0.346\beta - 0.0033\beta^2,$$
$$(\beta \sim 90.55 - 0.0077\alpha^2 - 0.95\alpha), \qquad (6)$$

where both $\alpha$ and $\beta$ are given in degrees, and the domain of alpha is restricted to [0,45]. In case of the stationary profiles of the experiment we measured $58° < \beta < 62°$ resulting in $24.7° < \alpha < 27.7°$.

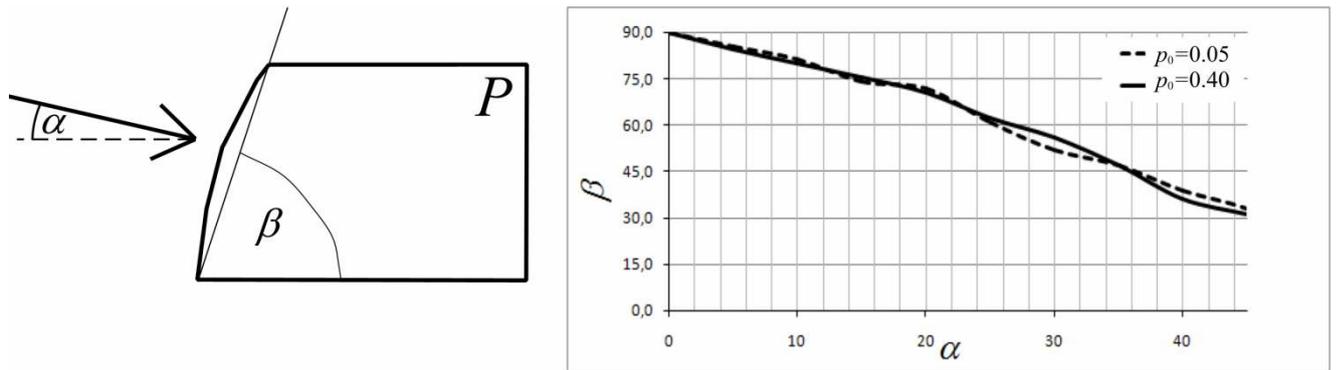

**Fig. 3** The strong coupling between $\alpha$ (slope of the incoming trajectory) and the average slope $\beta$ of the profile. Observe the almost linear $\alpha$-$\beta$ trend which appears to be extremely robust with respect to variations of $p_0$.

Rows in Figure 4 illustrate the effect of different average directions of impacting surfaces, the middle row corresponding approximately to the current experiment. Columns illustrate the effect of grain size, the last column corresponding approximately to the current experiment.

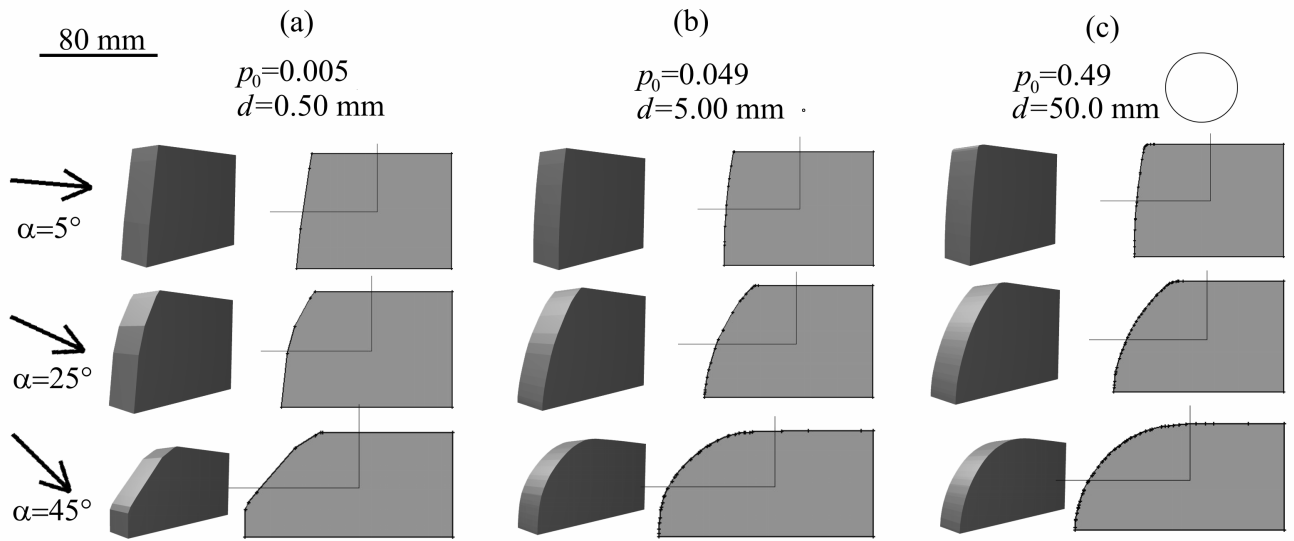

**Fig. 4** Qualitative summary of stationary profiles as control parameter $p_0$ and impact angle α are varied. Rendered 3D image is shown next to side view. Left column corresponds to sand blasting (grain diameter 0.5mm), resulting in rugged profiles with marked edges, right column to wear by large (50mm) impactors, corresponding to smooth surfaces. Impact angle α is governed by the wear mechanism: high impact angles (middle and lower rows) correspond to bouncing individual trajectories, low impact angle (upper row) to floating impactors. Observe wear of upper horizontal plane in case of high impact angle.

The time evolution computed in our model agrees well with the flume experiments. As we mentioned above, the expected value of the abraded area $\Delta$ plays the role of time. Figure 5 shows the time evolution of a profile computed with parameters matching closest the experimental values: $p_0$=0.40, α =26° and $\Delta$=0.01. We can observe the emergence of stationary profiles both in experiment and computation, good agreement of geometric shapes and good agreement of time evolution.

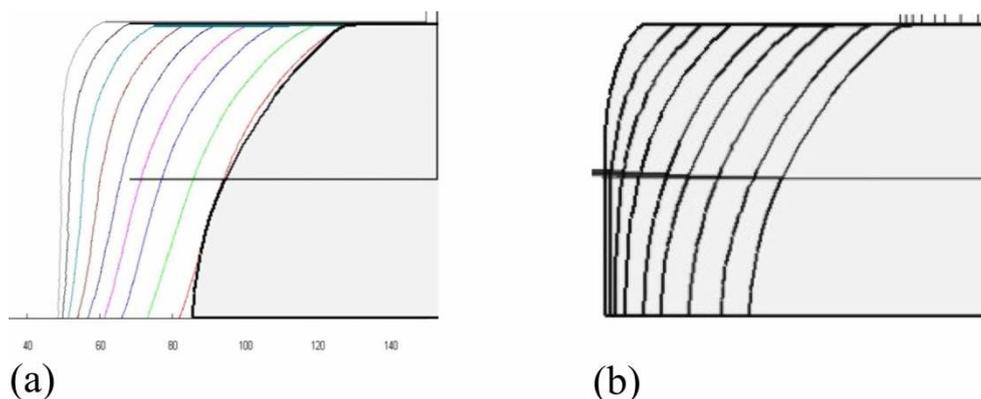

**Fig. 5** Comparison between computations and experiment: bedrock abrasion. Evolution of profile in experiment (a) and simulation (b). Observe similar evolution and appearance of stationary profiles. Each intermediate profile has been selected according to identical timing in computation and experiment, showing good agreement in the time evolution. Next to the last (stationary) profile in the experiment we plotted the last (stationary) profile of the computation for better comparison.

A similar comparison can be done with a much older experiment from (Schoewe 1932). This work aimed to investigate the morphology of ventifacts due to sand blasting. Sand particles used in that experiment resulted in a polygonal profile; this agrees well with the typical outcome of our model for small values of $p_0$. We found the best agreement to Figure 4 of (Schoewe 1932) at $p_0$=0.035, $\alpha$=30° and $\Delta$=0.002 (Fig. 6). In this case approximately 400 steps in the model describe 15 minutes of abrasion.

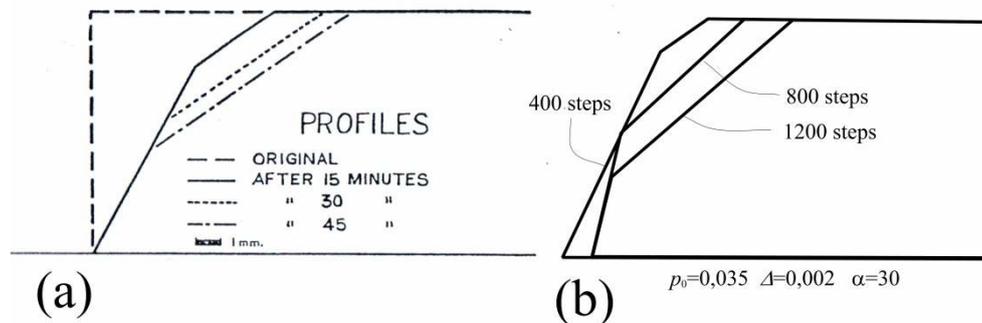

**Fig. 6** Comparison between computations and experiment: sand blasting. Evolution of profile in experiment (Schoewe 1932) (a) and simulation (b). In the computation, appearance of the steady state angle of the lower part of the profile is slower (it appears later, than 400 steps, i.e., 15 minutes), however, the full size of the experimental specimen (i.e., the parts not indicated in the figure) were not published in the original paper and this might explain the difference: in the model we restricted the size of the pebble to the visible part.

## 4. Conclusions

In this paper we presented a discrete, random model to describe bedrock abrasion. Our computations showed the robust emergence of steady state shapes, both the geometry and the time evolution of which agree with laboratory experiments and field observations. We found that small abrading grains result in polygonal profiles with edges and flat faces whereas large grains create smooth. convex profiles. These results confirm the arguments of (Wilson 2009) about the key role of the bedload in the erosion of bedrock bedforms.

## Acknowledgements

András Á. Sipos and Gábor Domokos have been supported by OTKA Grant #72146. Andy Wilson was supported by a NERC studentship held at the Department of Earth Sciences of the University of Cambridge. We are grateful to Jerome Lave for access to his flume for bedload abrasion experiments.